# Recommendation Algorithms that Increase Access to Influencers in a Network

By: Naisha Agarwal

Mentor: Professor Julian Shun

# Abstract

This paper introduces and studies the novel problem of improving fairness in networks as defined by access to influencers. It measures this fairness by the proximity of nodes to influencers and normalizes the value through a synthetic power-law graph of the same size using the Barabasi-Albert algorithm. It also proposes novel node recommendation algorithms to increase the fairness. The algorithm works by recommending nodes using standard node recommendation algorithms that are based on the number of triangles between the source and target node with probability *P*. With probability 1-*P*, it introduces weak ties and diversity in the network by recommending nodes using an importance sampling algorithm. This sampling algorithm is based on a polynomial function of the degree of the target node and its distance from the influencer set. Through extensive simulations on three real-world network data sets and comparing seven different algorithms, I show that the algorithm that adds diversity with probability proportional to the square of the ratio of the degree of the target node to distance to influencer achieves the best fairness. I also study and show the robustness of the algorithm to different parameter choices and provide insights on when to use the different importance sampling methods based on the structure of the network. The generalization of the proposed method to disconnected graphs is also discussed and illustrated.

# 1. Introduction

Networks are ubiquitous and influence our lives in significant ways. Often these networks have a small set of nodes that are highly influential and having access to these nodes is a significant advantage. For instance, on a social media site like LinkedIn, assuming everything else in terms of skills and other characteristics to be the same, access to decision makers at large corporations can increase the likelihood of obtaining jobs and other opportunities like sales. On Twitter and Facebook, access to influencers can help businesses market messages more easily and extensively. In academic networks, access to influential researchers can increase the likelihood of receiving a favorable recommendation. When deciding to build a new road, easy access to the more central roads can lead to better commerce and trade for the local township.

To ensure a level playing field for all nodes in a network, we must ensure nodes in a graph have similar access to influencers as much as possible. However, in many real world applications, new edges are formed via recommendation systems that are based on network attributes like triangle closings (number of mutual friends) and node characteristics (ex. demographics, industry, school, interests, etc). While these systems are great at optimizing the number of edges that are formed in the network over time, they create unintentional biases in the system where some nodes have an unfair advantage of gaining more access to opportunities and utilities through influencers. The algorithms that power these recommendation systems further exacerbate the situation due to the attributes they rely on. For instance, on LinkedIn, we ideally want a system where professionals with the same skills and qualifications have equal access to opportunities. However, due to the network bias that is created, this is hardly the case. Professionals who have large personal networks get connected to influential people more easily than those who are new to LinkedIn or those who do not have a strong network. Such network effects eventually create bias in who has access to opportunities in the future.

There is a need to change the algorithms powering these recommendations to reduce this network bias such that everyone is allowed equal access to opportunities based on what they know and not who they know. I show that it is indeed possible to do so by introducing some randomness in a controlled fashion to existing node recommendation algorithms. One way to view the randomness is by thinking of it as the introduction of some weak ties in the graph that increases access to influential nodes for more nodes in the network. For instance, seminal research by Mark Granovetter (Granovetter, 1973) showed that introducing weak ties in a network of professionals increases the likelihood of a node finding a job. To the best of my knowledge, a systematic study of introducing this diversification into node recommendation algorithms in the context of networks to increase fairness has not been studied before.

Since I don't have access to real recommendation systems, this paper uses simulation on real network datasets from various applications to study the problem and illustrate the proposed methodology.

It is important to study the problem of fairness due to the societal conditions that prevail in the world today. There are people who are otherwise qualified but who are not necessarily obtaining the opportunities they deserve because they may not always have the right connections. Social media plays a significant role in increasing unfairness and a large part of that is due to the algorithms that power their recommendation systems. These systems often aim at optimizing these algorithms to increase engagement and revenue via advertising through increasing the volume of connections. This unintentionally ends up increasing unfairness for nodes, especially those who do not know influencers and who are not part of social circles that include them. This research investigates one way of mitigating this bias in recommendation algorithms. By doing so, I hope people will not be penalized based on *who* they know and will get rewarded purely based on *what* they know. This will make our society much more equitable even as the influence of social media and digitization keeps increasing.

## 2. Background

I will use graphs and networks interchangeably in this paper. Graphs are used to represent many different scenarios, including social networks, road networks, power grid networks, among others.

To measure the influence of a node on the network, I use *betweenness centrality* as the measure of centrality. It measures the proportion of times a node acts as a bridge along the shortest path between two other nodes (Brandes, 2001). The equation for betweenness centrality for a node *v* is as follows:

$$c(v) = \sum_{s \neq v \neq t \in V} \frac{\sigma st(v)}{\sigma st}$$

where $\sigma_{st}$ is the total number of shortest paths from a source node *s* to a target node *t* and $\sigma_{st}(v)$ is the number of those paths that pass through *v* (Freeman, 1977). In other words, it is the proportion of shortest paths in the network that involve *v*.

I will now describe the novel fairness measure introduced in this paper. Fairness is defined to be the minimum shortest path to the influencer set. Using Dijkstra's algorithm (Dijkstra, 1959), the fairness of a node can be measured to be:

$$fairness(v) = min\{s \text{ in } S\} dist(v, s)$$

where *S* is the influencer set and *v* is a node in the graph.

The fairness of a graph is defined to be the average of the top *t%* of node fairness values. This can be modeled by the equation:

$$fairness(G) = top\ t\%\ \{v\ in\ V\ and\ fairness(v) \neq \infty\}\ fairness(v)$$

where *G* is the graph in question and *V* is the set of vertices in the graph not including influencers. I assume the graph *G* is connected; otherwise, the fairness would have been returned as infinity. Fairness can still be studied for disconnected graphs, but I describe this case separately in the technical approach section.

To normalize the fairness measure above to make it independent of the number of nodes and edges in the observed graph (scale free), I compute the fairness of a random graph having the same number of nodes and edges as the observed graph. The two methods I used in this paper to produce random graphs are the *Erdos-Renyi* graph and the *Barabasi-Albert* graph.

The *Erdos-Renyi* graph connects nodes at random with every edge being included in the graph with probability *p*. The probability for generating a graph with *n* nodes and *m* edges can be modeled by:

$$p^m (1-p)^{\left(\binom{n}{2} - m\right)}$$

where as probability *p* increases, it becomes more likely to add more edges to the given graph (Erdos & Renyi, 1959).

The *Barabasi-Albert* graph begins with an initial connected network of *n* nodes, where new nodes are added to existing nodes with probability $p_i$ where $p_i$ is the probability that a new node is connected to node *i*. This is modeled through the equation:

$$p_i = k_i / \Sigma_j k_j$$

where $k_i$ is the degree of node *i* and the baseline sum is made over all pre-existing nodes *j* (Barabasi & Albert, 1999).

# 3. Technical Approach

I will assume the given graph has nodes with the same attributes and the edges represent the connections among those nodes. In other words, the likelihood of any pair of nodes connecting is exactly the same in terms of attributes. They differ only because of graph structure (e.g., due to differences in the number of mutual edges).

There are two main parts to the technical approach I propose in this paper:

    1) finding an appropriate measure for fairness; and

    2) finding an approach to modify the node recommendation algorithms that are used in practice such that the modification introduces more fairness as measured by the index above.

## 3.1 - Finding an Appropriate Measure for Fairness

This part has three main methods:

    1) finding the influencers in the graph using graph theoretic measures;

    2) finding the fairness measure of each node and of the overall graph; and

    3) normalizing the graph fairness measure using a random graph to obtain a fairness index that makes it possible to compare fairness after adjusting for the varying number of edges and nodes in the graph.

*3.1.1 - Finding Influencers*

    I find influencers by calculating the betweenness centrality of every node in the graph and take the top *k* percent of nodes as the influencers.

*3.1.2- Finding The Fairness Measure*

    I first find the fairness of a single node and then aggregate it to compute the fairness of the entire network. If a path exists from the node to the influencer set, I define node fairness as the shortest path of the node to the influencer set (computed using the Djikstra's algorithm). If no path exists from a node to the influencer set, individual node fairness is not defined.

*3.1.2.2- Getting Fairness of Total Graph*

    For a connected graph, network fairness is defined as the average of the top *t* percent of node fairnesses. If the graph is disconnected, some nodes may not have any path to the influencer set. In this case, I define fairness as the proportion of nodes that cannot reach the influencer path. The higher the proportion, the more unfair the graph is.

*3.1.3- Calculating the Fairness Index*

Since graphs have different numbers of nodes and edges, it is important to normalize the fairness values to ensure it is scale free for connected graphs (the fairness measure for disconnected graphs is already scale free). I do so by using two random graphs: Erdos-Renyi and Barabasi-Albert.

## 3.2 Recommendation Algorithms to Increase the Fairness Index in a Social Network

The main idea of the proposed recommendation algorithm is motivated by a similar idea in the Watts-Strogatz model (Watts & Strogatz, 1998) in the context of the small-world experiment and decentralized search. The high level mathematical model behind this approach is as follows. For a given node *n*, with some probability *P*, the proposed algorithm recommends a node based on the number of mutual friends between *n* and the candidate node (the larger the mutual friends, the more likely it is to recommend the node). With probability (1-*P*), it adds some randomness and creates a weak tie by forming a new edge with a randomly selected target node. I also introduce a modification where the weak tie is selected via an importance sampling algorithm that is proportional to some function based on degree and distance to the influencer set of the candidate node. More details are provided below.

*3.2.1 - Details*

For a given node *n*, consider the array $M[n] = \{(m, i(m), w(i(m))\}$. Here, *m* denotes the target candidate nodes that *n* is not yet connected to. For each node *m*, $i(m)$ denotes the number of mutual friends (edges) between *n* and *m* and $w(i(m))$ denotes the weight of *m*. The influencer set is not included in *M[n]* to avoid trivial solutions.

*Calculate w(i(m)) values*

The weight *w(i(m))* should be an increasing function of *i(m)* because the more mutual friends there are between *n* and *m*, the more likely they are to connect. Therefore, they should be recommended with a higher probability. One popular *w* function that is used is the sigmoid function $q(i) = 1/(1 + e^{a-i})$ where *a* is a constant. Since $q(0) = 1/(1 + e^a)$ is a constant, I can choose $w(i) = q(i)/q(0) = (e^{-a} + 1)/(e^{-a} + e^{-i}) \approx e^i$. I make this approximation assuming *a* is a very large number. Hence, $w(i) = e^i$.

Now consider another array for a given node *n* as $R[n] = \{m, D(m), d(m), wr(a, b))\}$. Here, *D(m)* is the degree of *m*, *d(m)* is the minimum distance of *m* to the influencer set, *wr(a,b)* is the weight function written as $wr(a, b) = D(m)^a / d(m)^b$, and *a* and *b* are non-negative constants. The reasoning behind choosing this weight function is described later below.

*Algorithm Details*

With probability *P*, the algorithm recommends a node for a given visit node *n* by sampling a node from *M* with weights proportional to *w(i)'s*. With probability *(1-P)*, it recommends a node by sampling from *R* with weights proportional to *wr(a,b)*. Note that *wr(0,0)* = 1. Just like in the Watts-Strogatz model, with probability *P*, the new algorithm recommends a node that is close to *n*'s neighborhood and with probability *(1-P)*, it selects a node at random (assuming *a=b=0*). These random nodes shorten the path to the influencer set and create more fairness.

For simplicity and efficiency in the simulations, the algorithm always connects to the recommended node. In practice, this does not happen and recommendations have different connection probabilities that are not constant. For instance, the connection probability to a node with zero mutual friends is typically lower than connecting to a node with ten mutual friends. Since the goal of this paper is only to study the impact of adding diversity to fairness, this assumption can be made without loss of generality. In the end, what really matters is the budget in terms of adding nodes at random and the fairness index it produces.

Like in the Watts-Strogatz model, diversity can be introduced in a controlled manner. I use two intuitions for this:

    1) Sampling higher degree nodes with higher weights is helpful since they are likely to create more paths to the influencer set; and

    2) Given two nodes that have the same high degree, the one that is closer to the influencer set is more likely to increase fairness.

However, it is not clear what is better: a very high degree node far away from the influencer set (e.g., a node with a degree of 100 that is a distance of 3 from an influencer) versus a less high degree node closer to the influencer set (e.g., a node with a degree of 50 that is a distance 1 from an influencer). In the simulations, I test multiple *a* and *b* values, which will be described in further detail in the experiments section.

This algorithm is illustrated below in pseudocode:

```
Parameters: N (number of visits) , a, b, P, inf (influencer array), G (graph)

1   g_new = G
2   nvisits = 0
3   while(nvisits < N):
4       select random node n
5       generate M(n)
6       generate R(n)
7       with probability P:
8           sample an edge from M
9       else with probability 1-P:
10          sample an edge from R
11      g_new = g_new ∪ new edge
12      nvisits+=1

function generateM(n):
1   get nodes that n is not connected to (excluding influencers)
2   compute mutual friends with n for each node
3   compute weights for each node using e^i where i is the number of mutual friends
function generateR(n):
1   get nodes that n is not connected to (excluding influencers)
2   compute degree and the minimum distance to an influencer of each node
3   sample with weights equal to D^a/d^b
```

To get more intuition into the algorithm, I compute the distance between $w(i)$ (the mutual friends algorithm) and $wr(a,b)$ using the Kullback-Leibler distance function. The distance is given by

$$KL(w, wr\ (a=0, b=0)) = A\sum_{k=1}^{K} f(k)ke^k + C$$

where $k$ denotes the number of mutual friends, $K$ is the maximum value of mutual friends there are in the data, $f(k)$ denotes the frequency of target nodes that have $k$ mutual friends with the source node, and $A$ and $C$ are constants. I assume $a=b=0$ for comparing the distance with a completely random distribution as the baseline. As can be seen above, the distance from the random distribution increases when the source node has a higher concentration of target nodes with a large number of mutual friends. These are the scenarios where the randomization is breaking the "rich gets richer" characteristics of the usual recommendation algorithm and creating more fairness. A detailed derivation is given below.

*Deriving the Kullback-Leibler Equation*

For a given source node *n*, let *m* subscript the target nodes the source node can connect to and let *P* and *Q* be the probability distributions for weights *w* and *wr* (*a*=*b*=0) respectively. Hence *P* = *Aw,* and *Q*=*B(wr),* where *A* and *B* are normalizing constants. Hence,

$$KL(P,Q) = \sum_m P(m)\log(P(m)/Q(m))$$

Since $w(m)=e^{i(m)}$ and *wr(a=b=0)* = 1, where *i(m)* is the number of mutual friends between *n* and target node *m*, the expression simplifies to:

$$KL(P,Q) = \sum_m A e^{i(m)}\log(A e^{i(m)}/B*1)$$

$$KL(P,Q) = A\sum_m i(m)e^{i(m)} + C; \text{ where } C \text{ is a constant.}$$

Let *k* be the number of mutual friends and f(*k*) denote the frequency of target nodes with *k* mutual friends with the source node *n*. Doing a frequency tabulation of $i(m)e^{i(m)}$ over all target nodes, the expression becomes:

$$KL(P,Q) = A\sum_{k=1} f(k) k e^k p(k) + C.$$

# 4. Experiments

I conducted simulations on multiple datasets to show the efficacy of the proposed methods: a Facebook social circles graph (McAuley & Leskovec, 2012), a LastFM graph (Rozemberczki & Sarkar, 2020), and a road graph (Rossi & Ahmed, 2015). The first two datasets are connected graphs while the last dataset is disconnected. Each of these data-sets has a unique network structure and provides complementary insights to the methodology in this paper. I also conduct a robustness study for the choice of an influencer set.

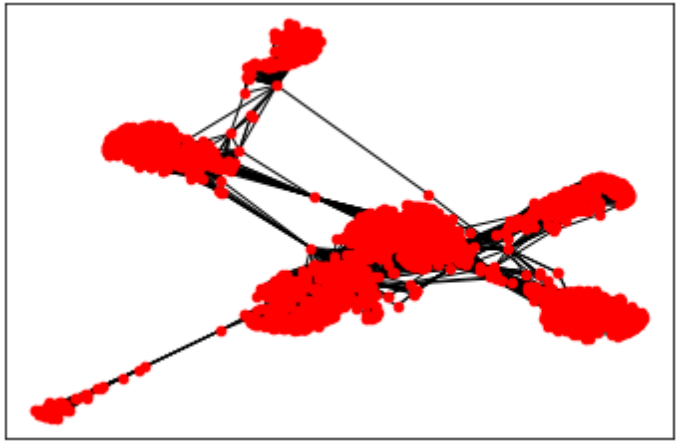

**Figure 1.** *Visualization of Facebook network using NetworkX draw feature*

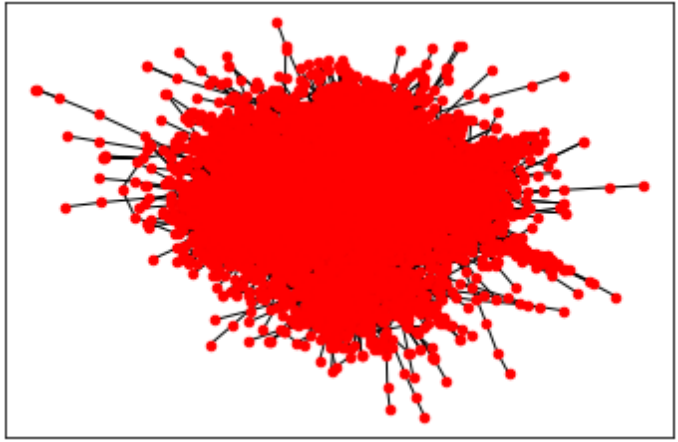

**Figure 2.** *Visualization of LastFM network using NetworkX draw feature*

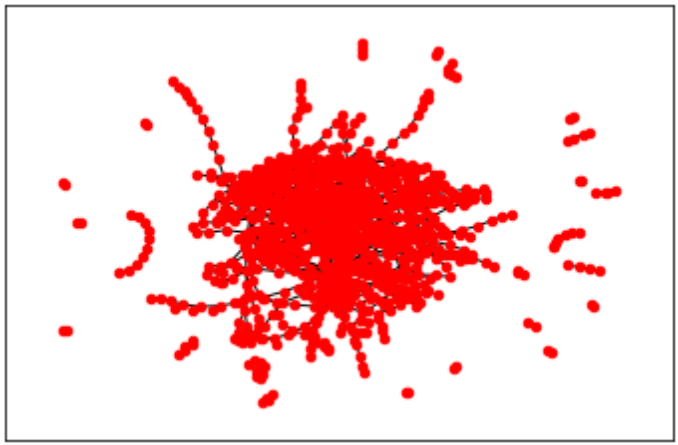

**Figure 3.** *Visualization of Road Network using NetworkX draw feature*

From Figure 1, it is apparent that in the Facebook data there are several dense communities loosely connected by some weak ties. From Figure 2, it is shown that there is one central community with some nodes on the periphery for the LastFM data. Figure 3 shows the road network as a sparse and disconnected graph.

**4.1- Experimental Setup**

All random graphs and visualizations were generated using python module NetworkX v2.5, and all other graphs were created with Python module Matplotlib v3.4.2.

For each data set, I ran the $D^a/d^b$ algorithm for the following pairs of *a* and *b:*

- *a* = 0, *b* =0 (completely random)
- *a* = 1, *b*= 0 (proportional to degree)
- *a* = 1, b = 1 (proportional to degree/distance (*D/d*) )
- *a* = ½, *b*= ½ (proportional to $\sqrt{D/d}$)
- *a* = 0, *b*= 2 (proportional to $1/d^2$)
- *a* = 2, *b*=2 (proportional to $(D/d)^2$)
- *a* = 1, *b*=2 (proportional to $D/d^2$)

I also calculated the fairness index of nodes by using the Barabasi-Albert random graph since it converges faster than the Erdos-Renyi random graph.

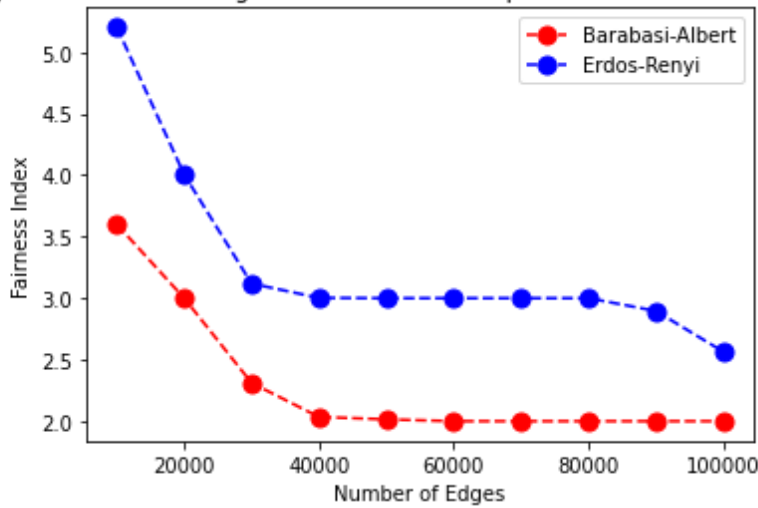

**Figure 4.** *Fairness Convergence for Random Graphs for Different Amounts of Edges*

As can be seen through Figure 4, the Barabasi-Albert random graph converges faster than the Erdos-Renyi graph, and therefore provides a better normalizing value for calculating the fairness index.

**4.2 - Facebook Data**

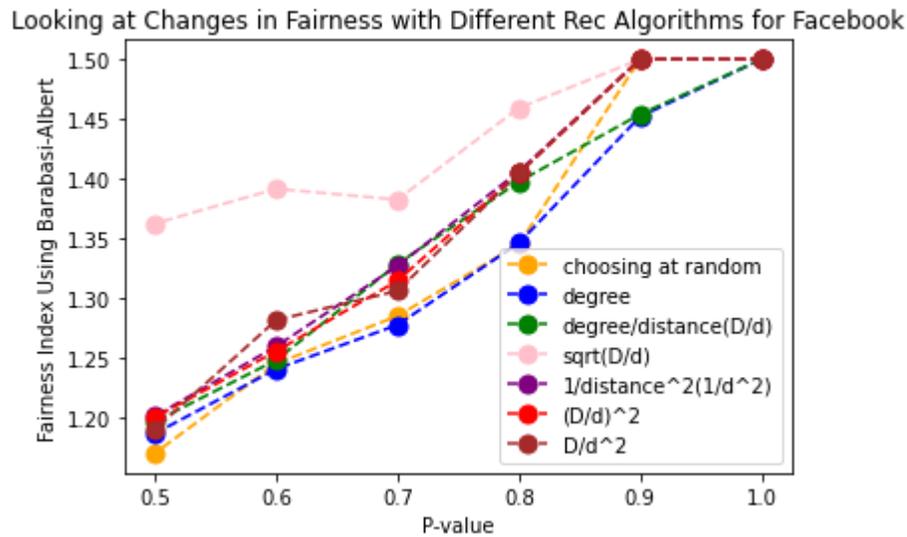

**Figure 5.** *Changes in Fairness with Different Recommendation Algorithms for Facebook (Lower Fairness Index means More Fairness)*

In each of these graphs, the lower the fairness index, the better the fairness in the graph since there are shorter paths to the influencer relative to the Barabasi-Albert graph. Therefore, in Figure 5, it can be observed that degree is the best algorithm for the Facebook graph as it is the consistently lowest line on the graph. $D/d^2$ and $(D/d)^2$ are close seconds.

It can also be observed that each of the algorithms have the same value for $P = 1$. This is because the algorithms at that point are completely based on mutual friends and not randomness. When decreasing $P$, meaning there is a higher chance of randomness in the network, the fairness index is decreasing as expected.

To investigate why degree was the best algorithm for the Facebook graph, I made a scatterplot comparing degree and distance to the influencer set.

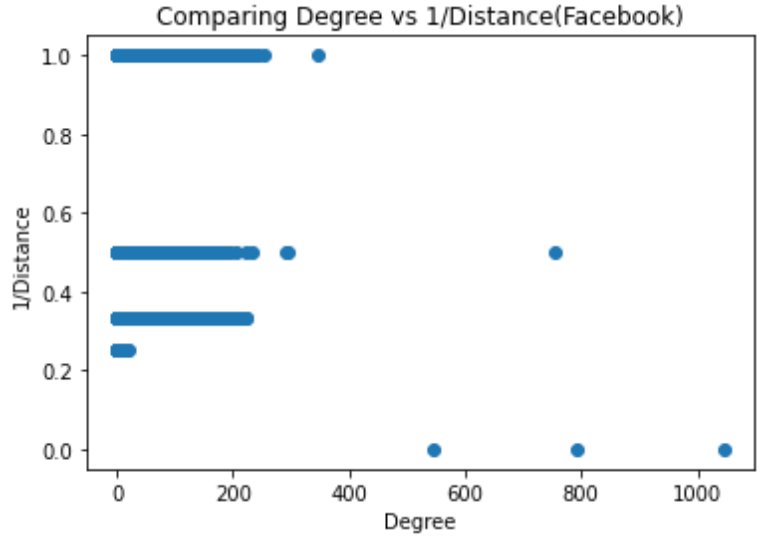

**Figure 6.** *Scatterplot of Facebook Data with Degree on x-axis and 1/Distance on y-axis*

From Figure 6, it can be seen how very high degree nodes are not close to the influencers and hence *D/d* selects them with very low probability, hurting the fairness performance of *D/d*. On the other hand though, there are a significant number of nodes with high-degrees and hence selecting based on just degree leads to the creation of more paths to influencers.

### 4.3 - LastFM Data

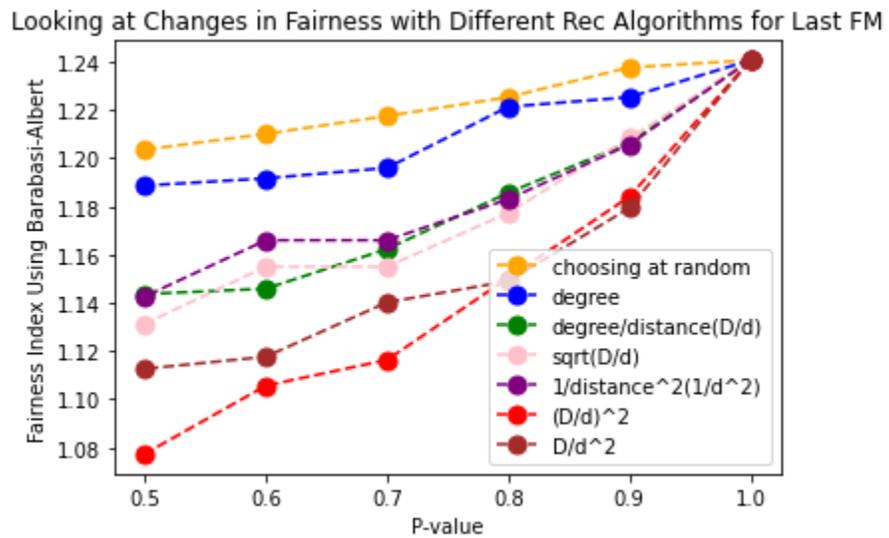

**Figure 7.** *Changes in Fairness with Different Recommendation Algorithms for Last FM (Lower Fairness Index means More Fairness)*

From Figure 7, it is observed that $(D/d)^2$ is clearly the best algorithm for the LastFM data because it is the lowest line on the graph. To investigate why, I made a scatterplot comparing degree and distance.

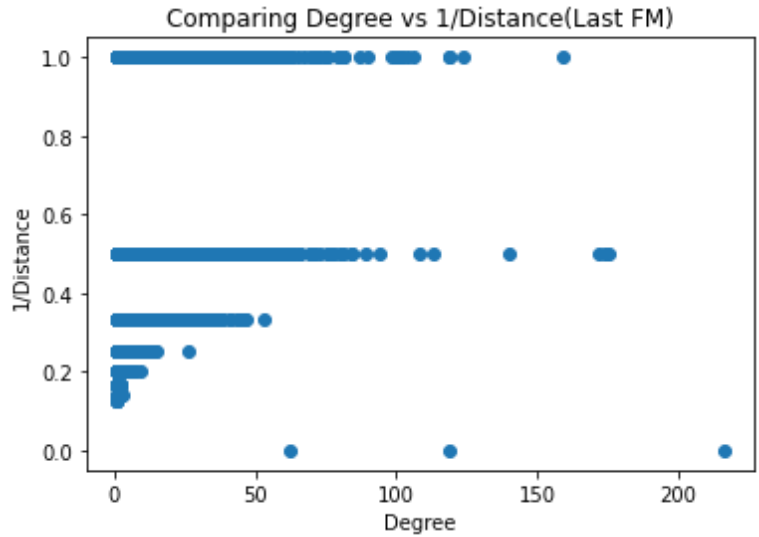

**Figure 8.** *Scatterplot of LastFM Data with Degree on X-axis and 1/Distance on y-axis*

In Figure 8, it can be seen how there are some high degree nodes that are farther away from the influencers while others are closer. Adding randomness just proportional to degree would select both nodes with that degree with the same probability regardless of their distance to the influencer. This is why selecting randomness proportional to $D/d$ is more accurate because it penalizes nodes that are far away from influencers despite having a high-degree.

The reason why $(D/d)^2$ seems to be the best algorithm is because it looks like the higher the power of $D/d$, the more the function penalizes the nodes that are far away from influencers, and the better the fairness value it returns.

## 4.4 - Studying the Impact of Changing Influencers on the Fairness Index

I manually changed the influencers of the LastFM graph to show the robustness of this algorithm. The new influencers had very low betweenness centralities, representative of situations where influencers who don't have a big social media presence are still influential in the real world.

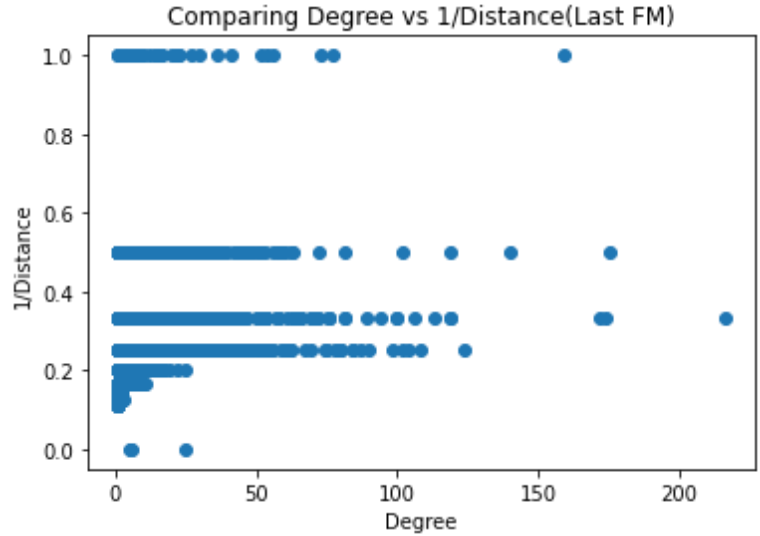

**Figure 9.** *Scatterplot of LastFM Data with Manually Changed Influencers*

From Figure 9, it can be seen how even with a change in influencers and a drop in fairness, the same pattern emerges where there are high-degree nodes both close to influencers and far away.

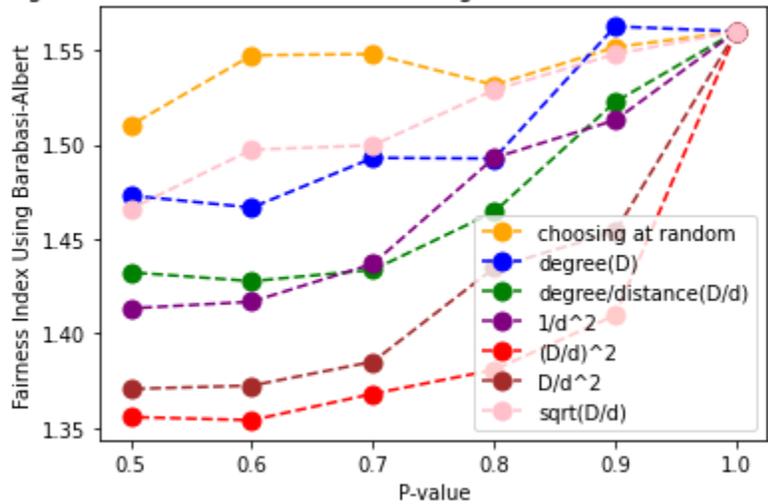

**Figure 10.** *Looking at Different Recommendation Algorithms with a Change in Influencers (Lower Fairness Index means More Fairness)*

Figure 10 further proves the points made from Figure 9, where $(D/d)^2$ is still clearly the best algorithm.

## 4.5 - Road Network Data

Unlike previous data sets examined, it is apparent from Figure 3 that the road network has many disconnected components, which drastically affects the results of the different recommendation algorithms proposed in this paper. Therefore instead of measuring fairness as the minimum distance to an influencer, I looked at the proportion of nodes that didn't have a path to the influencer as a function of different *P*-values.

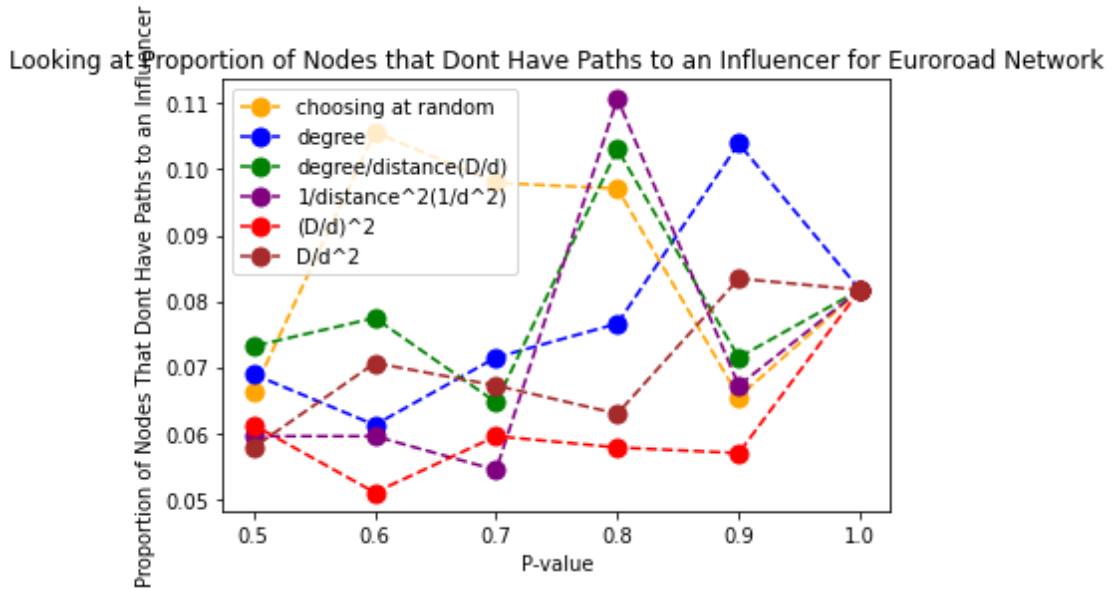

**Figure 11.** *Looking at Different Recommendation Algorithms for a Road Network*

From Figure 11, it can be seen how increasing the amount of randomness in the network overall decreased the percentage of nodes that didn't have a path to the influencer, and hence increased fairness. This data set shows how the algorithm can work with not just connected graphs, but disconnected ones as well.

## 4.6 - Summary of Experimental Results

Overall, the main findings were:
1) As the amount of randomness in the network is increased (decreasing *P*-value), the overall fairness in the network in terms of access to opportunity increases.
2) Overall, it seems like $(D/d)^2$ is the best recommendation algorithm to increase fairness in the network, although it is specific to the data-set at hand.

3) There is robustness in the algorithm with the number of influencers and whether the graph is connected or not; although the overall fairness deteriorates, it still follows the same pattern.

# 5. Related Work

Social networks are not always fair. Whether it be Facebook, Instagram, LinkedIn, a sales network, or others, there are always some users that have an advantage over others simply because of who they are connected to. In fact, there has been a recent study done by the MIT Technology Review on Facebook and how their ad algorithm discriminates by race by recommending low-paying jobs more frequently to Mexicans and Hispanics than Caucasians (Hao, 2019). There have also been studies on the bias in LinkedIn machine learning models and how the recommendation systems that LinkedIn uses are inherently biased (Saint-Jacques, G., Sepehri, A., Perisic, I., & Li, N, 2020).

The above are only some documented examples of unfairness in networks; there are many countless others that may exist. Although it is hard to completely eradicate all kinds of unfairness in networks, we can certainly reduce them, especially those that arise due to algorithms that are used to form new edges through recommendations. Not much research has been done in this area, but this is an important problem. This paper attempts to find ways of reducing unfairness in a network by modifying the underlying recommendation algorithms.

The mathematical model used to modify the recommendation algorithm in this paper is related to the model studied by Watts and Strogatz (W-S). In their paper, W-S present a network where there are many links that form due to homophily and triadic closure, while also presenting weak long-range ties to properly simulate the small-world phenomenon. It was the first model that showed how a small amount of randomness can lead to more connections in a social network and reduce the diameter of the network. However, the main motivation of W-S was to perform efficient decentralized search, finding the best way to efficiently transmit a message to a target node from a source node using the existing network. The focus in this paper is to recommend new nodes to existing nodes and in doing so ensure equal access to influencers (a set of target nodes). The mathematical model of introducing controlled diversity is also different from the Watts- Strogatz model as I use a function of both degree of a node and its distance to the influencer set and show that this function leads to significant improvements in fairness.

Another seminal paper that studied the problem of increasing weak ties in networks was *The Theory of Weak Ties* (Granovetter, 1973). Granovetter did a survey in Boston and empirically showed that adding weak ties increased the likelihood of nodes getting connected to influential people (job posters) and finding a job. However, his paper did not study a method to improve recommendations to increase such fairness; it only showed that the existence of such ties was effective.

Before we can increase fairness, it is important to have a principled method to measure it. There have been many solutions proposed to measure fairness in networks. These include the Atkinson Index (Saint-Jacques, G., Sepehri, A., Perisic, I., & Li, N, 2020) and using various other algorithms in networks (Mehrabi, Morstatter, Saxena, Lerman & Galstyan, 2019; Mehrabi, N., Morstatter, F., Saxena, N., Lerman, K., & Galstyan, A., 2019). However, the fairness measure proposed in this paper is different from previous ones; it measures the shortest distance to an influencer set in a network and finds the fairness of a graph by taking the average of the top $t$% of node fairnesses. While there has been research done on determining influencers in a network (Aman Ullah, Bin wang, Jinfang Sheng, Jun Long, Nasrullah Khan, 2021; J. Dai et al. 2019; B. Rozemberczki, R. Davies, R. Sarkar and C. Sutton, 2018), my approach goes a lot further. It calculates a scale free fairness measure using the influencer set, and improves the fairness in a network via novel node recommendation algorithms.

# 6. Conclusion

In this paper, a new way of measuring fairness was introduced. Fairness was defined to be the shortest distance from a node to any of the influencers in the graph, and a fairness index was calculated in respect to the Barabasi-Albert random graph. Through different recommendation algorithms, this paper showed how increasing the amount of randomness in a network increases fairness and gives users opportunities to connect with influencers. It also showed that overall $(D/d)^2$ is the best algorithm for increasing fairness and that this algorithm is robust with changes in influencers and whether a graph is connected or not.

In terms of future directions of this paper, the methods discussed can be scaled to larger graphs that are more representative of current social networks. One can also use other measures of centrality to identify the influencers in a network and see how that impacts the results.

The activeness of a user can also be taken into account when running simulations. Currently, I assume that all nodes are equally likely to visit the network in the simulations; however, in reality, it depends on how active an user is in the network that determines how often they visit. In general, nodes with higher degrees tend to visit more often. Additionally, in this paper I assume that the number of nodes in a graph is constant. However in reality, users are constantly joining and leaving social networks, which will affect the total fairness. This is something that can be studied in further research. One can also look at other importance sampling functions and see if any of those increase the fairness more than the $D^a/d^b$ algorithm.

Lastly, I want to run these experiments on real recommendation systems by working at a social media company. Such experiments would also help me understand the opportunity cost in terms of engagement loss and revenue of the algorithm as it introduces different amounts of fairness.